\documentclass[prd,twocolumn,showpacs,preprintnumbers,amsmath,amssymb,superscriptaddress,floatfix,nofootinbib]{revtex4-2}
\usepackage{graphicx}
\usepackage{amsmath}
\usepackage{amsfonts}
\usepackage{amssymb}
\usepackage{color}
\usepackage{multirow}
\usepackage[colorlinks, citecolor=blue,anchorcolor=red,menucolor=red,linkcolor=red,filecolor=red,runcolor=red,urlcolor=blue,frenchlinks=red]{hyperref}

\usepackage{subfigure}

\begin{document}    

\title{Theoretical study of scalar meson $a_0(1710)$ in the $\eta_c \to  {\bar{K}}^0K^+\pi^- $ reaction}

\date{\today}
\author{Yan Ding}
\affiliation{School of Physics and Microelectronics, Zhengzhou
	University, Zhengzhou, Henan 450001, China}

\author{Xiao-Hui Zhang}
\affiliation{School of Physics and Microelectronics, Zhengzhou
	University, Zhengzhou, Henan 450001, China}
 
\author{Meng-Yuan Dai}
\affiliation{School of Physics and Microelectronics, Zhengzhou
	University, Zhengzhou, Henan 450001, China}

\author{En Wang}\email{wangen@zzu.edu.cn}
\affiliation{School of Physics and Microelectronics, Zhengzhou
	University, Zhengzhou, Henan 450001, China}
\affiliation{Guangxi Key Laboratory of Nuclear Physics and Nuclear Technology, Guangxi Normal University, Guilin 541004, China}

\author{De-Min Li}\email{lidm@zzu.edu.cn}
\affiliation{School of Physics and Microelectronics, Zhengzhou
	University, Zhengzhou, Henan 450001, China}

\author{Li-Sheng Geng} \email{lisheng.geng@buaa.edu.cn}
\affiliation{School of Physics, Beihang University, Beijing 102206,
China}
\affiliation{Beijing Key Laboratory of Advanced Nuclear Materials and Physics, Beihang University, Beijing, 102206, China}
\affiliation{Peng Huanwu Collaborative Center for Research and Education, Beihang University, Beijing 100191, China}
\affiliation{Southern Center for Nuclear-Science Theory (SCNT), Institute of Modern Physics, Chinese Academy of Sciences, Huizhou 516000,  China}

\author{Ju-Jun Xie} \email{xiejujun@impcas.ac.cn}
\affiliation{Institute of Modern Physics, Chinese Academy of
Sciences, Lanzhou 730000, China} \affiliation{School of Nuclear
Sciences and Technology, University of Chinese Academy of Sciences,
Beijing 101408, China}
\affiliation{Southern Center for Nuclear-Science Theory (SCNT), Institute of Modern Physics, Chinese Academy of Sciences, Huizhou 516000, Guangdong Province, China}

\begin{abstract}

We investigate the process $\eta_c \to {\bar{K}}^0K^+\pi^-$ by taking into account the $S$-wave ${K^*\bar{K}^*}$ and $\rho\omega$ interactions within the unitary coupled-channel approach, where the scalar meson $a_0(1710)$ is dynamically generated. In addition, the contributions from the intermediate resonances $K_0^*(1430)^{-}\to {\bar{K}}^0\pi^- $ and $K_0^*(1430)^{0}\to K^+\pi^-$ are also considered. We find a significant dip structure around 1.8~GeV, associated to the $a_0(1710)$, in the ${{\bar{K}}^0K^+}$ invariant mass distribution,  and the clear peaks of the $K_0^*(1430)$ in the ${\bar{K}}^0\pi^-$ and $K^+\pi^-$ invariant mass distributions, consistent with the {\it BABAR} measurements. We further estimate the branching fractions $\mathcal{B}(\eta_c \to \bar{K}^{*0}K^{\ast+}\pi^-)= 5.5\times10^{-3}$ and $\mathcal{B}(\eta_c \to \omega\rho^+\pi^-)= 7.9\times10^{-3}$. Our predictions can be tested by the BESIII and Belle II experiments in the future.
\end{abstract}

\maketitle

\section{Introduction} \label{sec:Introduction}
 In 2021, the $\textit{BABAR}$ Collaboration  observed the scalar resonance $a_0(1710)$ in the $\pi^\pm\eta$ invariant mass spectrum of the process $\eta_c\to \eta\pi^+\pi^-$~\cite{BaBar:2021fkz}. In 2022, the BESIII Collaboration also found the  $a_0(1710)$ state in the $K_S^0K_S^0$ invariant mass spectrum of the process $D_s^+ \to K_S^0 K_S^0\pi^+$~\cite{BESIII:2021anf}, and in the $K_S^0K^+$ invariant mass spectrum of the process $D_s^+ \to K_S^0 K^+\pi^0$~\cite{BESIII:2022npc}. The experimental measurements of the mass and width of $a_0(1710)$ are tabulated in Table~\ref{tab:exp.mass and width}. One can see that there are some discrepancies between the measured masses.
 Note that in Ref.~\cite{BESIII:2021anf}, BESIII did not distinguish between the $a_0(1710)$ and $f_0(1710)$ in the process $D_s^+ \to K_S^0 K_S^0\pi^+$, and denoted the combined state as $S(1710)$, while in Ref.~\cite{BESIII:2022npc} the $a_0(1710)$ was renamed as $a_0(1817)$ because of the different fitted mass of this state.

It should be stressed that there have been many theoretical studies about the structure of the $a_0(1710)$ and its isospin partner $f_0(1710)$ from various perspectives~\cite{Nagahiro:2008bn,Branz:2009cv,Geng:2010kma,Xie:2014gla,MartinezTorres:2012du,Wang:2021jub,Garcia-Recio:2010enl,Garcia-Recio:2013uva,Close:2005vf,Gui:2012gx,Janowski:2014ppa,Fariborz:2015dou,Chen:2005mg}.  
For the $f_0(1710)$, although it is a well-established state according to the Review of Particle Physics (RPP)~\cite{ParticleDataGroup:2022pth}, there are still different interpretations of  its structure. In Ref.~\cite{Close:2005vf}, it was shown that the $f_0(1710)$ wave function contains a large $s\bar{s}$ component, while in Refs.~\cite{Gui:2012gx,Janowski:2014ppa,Fariborz:2015dou,Chen:2005mg}, it was regarded as a scalar glueball. In addition, the $f_0(1710)$ and $a_0(1710)$ states could be dynamically generated from the vector-vector interactions~\cite{Geng:2008gx,Du:2018gyn}, and this picture remains essentially the same when the pseudoscalar-pseudoscalar coupled-channels were taken into account~\cite{Wang:2022pin}. In Ref.~\cite{Wang:2017pxm}, one isovector  scalar state $a_0$ with a mass of 1744~MeV is also predicted in the approach of  Regge trajectories, which is roughly consistent with the experimental mass of the $a_0(1710)$.

\begin{table}[htbp]
	\begin{center}
		\caption{\label{tab:exp.mass and width} Experimental measurements on the mass and width of  the scalar state $a_0(1710)$. The first error is statistical and the second one is systematic. All values are in units of MeV.}
		
		\begin{tabular}{cccc}\hline\hline
			
			Experiment & $M_{a_0(1710)}$ \quad & $\Gamma_{a_0(1710)}$ \quad &Reference \\ \hline
			$\textit{BABAR}$  \quad  & $1704 \pm 5 \pm 2$ \quad  &\quad $110\pm 15\pm 11$ &~\cite{BaBar:2021fkz}  \quad       \\
			BESIII      \quad     & $1723 \pm 11 \pm 2$ \quad  &\quad $140 \pm 14 \pm 4$ &~\cite{BESIII:2021anf} \quad     \\
			BESIII     \quad    &  $1817\pm 8 \pm 20$ \quad  &\quad
			$97 \pm 22 \pm 15$ &~\cite{BESIII:2022npc} \quad    \\
			\hline\hline			
		\end{tabular}
	\end{center}
\end{table}

As shown in Table I,  the mass of the $a_0(1710)$ is not well determined experimentally. This can complicate the understanding of  the nature of the $a_0(1710)$. For instance, $a_0(1710)$ (or $a_0(1817)$) and $X(1812)$ have been explained as the $3^{3}P_0$ state by assuming $a_0(980)$ and $f_0(980)$ as $1^{3}P_0$ states~\cite{Guo:2022xqu}. Indeed, $X(1812)$ was observed in the process $J/\psi \to \gamma\phi\omega$ by the BESIII Collaboration~\cite{BES:2006vdb,BESIII:2012rtd}, and the enhancement near the $\phi\omega$ threshold, associated to $X(1812)$, could be described by the reflection of $f_0(1710)$, as discussed in Ref.~\cite{MartinezTorres:2012du}.
By regarding the $a_0(1710)$ as a $K^*\bar{K}^*$ molecular state, Refs.~\cite{Zhu:2022wzk,Zhu:2022guw,Dai:2021owu,Oset:2023hyt,Wang:2023aza}  have successfully described  the invariant mass distributions of the processes $D_s^+ \to K_S^0 K_S^0\pi^+$ and  $D_s^+ \to K_S^0 K^+\pi^0$ measured by the the BESIII Collaboration .

Since the peak positions of the $a_0(1710)$ in the $K\bar{K}$ invariant mass distributions of the processes $D_s^+\to K_S^0 K_S^0 \pi^+, K_S^0K^+\pi^0$ observed by the BESIII Collaboration are very close to the boundary region of the $K\bar{K}$ invariant mass, it is crucial to measure the properties of  the $a_0(1710)$ precisely in other processes with larger phase space~\cite{Wang:2023lia}.  Taking into account that the dominant decay channel of the $a_0(1710)$ is $K\bar{K}$ in the molecular picture~\cite{Geng:2008gx,Wang:2022pin}, we propose to search for this state in the process $\eta_c\to \bar{K}^0K^+\pi^-$.  Indeed, there have been some experimental studies of this process. In 2012, the BESIII Collaboration has measured the branching fraction $\mathcal{B}(\eta_c\to K_S^0K^{\pm}\pi^{\mp})=(2.60\pm0.29\pm0.34\pm0.25) \%$ via $\psi(3686)\to \pi^0 h_c, ~ h_c \to \gamma \eta_c$ with a sample of 106 million $\psi(3686)$ events~\cite{BESIII:2012urf}. In 2019, the BESIII Collaboration measured the branching fraction of this process $\mathcal{B}(\eta_c\to K_S^0K^{\pm}\pi^{\mp})=(2.60\pm0.21\pm0.20) \%$ via $e^+e^-\to \pi^+\pi^- h_c,~ h_c \to \gamma \eta_c$ with the data samples collected at $\sqrt{s}=4.23$, $4.26$, $4.36$, and $4.42$~GeV~\cite{BESIII:2019eyx}. In addition, the {\it BABAR} Collaboration has observed this process in the $\gamma\gamma^*\to\eta_c\to K_S^0K^{\pm}\pi^{\mp}$~\cite{BaBar:2010siw,BaBar:2015kii}, and the measured $K_S^0K^+$ mass  spectrum shows some structure in the region of 1.7$\sim$1.8~GeV, which could  hint at the existence of the $a_0(1710)$, as we show in this work.

Based on the $\textit{BABAR}$ data~\cite{BaBar:2010siw,BaBar:2015kii}, we will investigate the process $\eta_c \to \bar{K}^0K^+\pi^-$. In addition to the contribution from the scalar resonance $a_0(1710)$, we also take into account  the contribution from the intermediate resonance $K_0^*(1430)$, which plays an important role in this process according to Refs.~\cite{BaBar:2010siw,BaBar:2015kii}.

The paper is organized as follows. In Sec.~\ref{sec:Formalism}, we present the
theoretical formalism of the $\eta_c \to \bar{K}^0 K^+\pi^-$ decay, and
in Sec.~\ref{sec:Results}, we show our numerical results and discussions,
followed by a short summary in the last section.

\section{Formalism} \label{sec:Formalism}

First in Subsect.~\ref{subsec:2A} we  present the theoretical formalism for the process $\eta_c \to \bar{K}^0 K^+\pi^-$ via the $K^*\bar{K}^*$ and $\omega\rho$ final state interactions, which dynamically generate the scalar resonance $a_0(1710)$ . Next, we show the formalism for the process of $\eta_c \to K_0^*(1430)^- K^{+} [K_0^*(1430)^0 \bar{K}^{0}]$  with $K_0^*(1430)^- \to K^0 \pi^-$ [$K_0^*(1430)^{0} \to K^+ \pi^-$] in Subsect.~\ref{Subsec:2B}. Finally, the formalism for the double differential widths of the process $\eta_c \to \bar{K}^0 K^+\pi^-$ is given in Subsect.~\ref{Subsec:2C}.

\subsection{Mechanism of $\eta_c \to  (\bar{K}^{*0}K^{\ast+} /\omega \rho^+)  \pi^- \to \bar{K}^0K^+\pi^-$}\label{subsec:2A}

\begin{figure}[tbhp]
    \centering
        \includegraphics[scale=0.8]{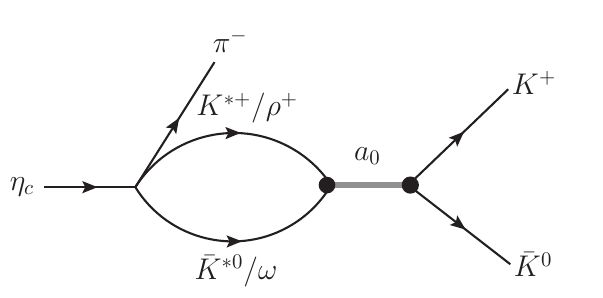}
    \caption{Diagram for the process $\eta_c \to \left(\bar{K}^{*0}K^{\ast+}/ \omega \rho^+\right) \pi^- \to a_0(1710)^+ \pi^-\to  \bar{K}^0K^+\pi^-$.}
  \label{fig:hadronFSI}
\end{figure}

With the assumption that the $\eta_c$ is a singlet of SU(3), and $a_0(1710)$ is a vector-vector molecular state~\cite{Geng:2008gx,Du:2018gyn}, one needs to first produce the vector-vector pairs in the $\eta_c$ decay. Considering that this process has a $\pi^-$ in the final states, we introduce one combination mode of $\textless VVP\textgreater$ in the primary vertex~\cite{Ikeno:2019grj,Jiang:2019ijx}, where $V$ and $P$ are the SU(3) vector and pseudoscalar matrices respectively~\cite{Ikeno:2019grj,Jiang:2019ijx,Zhang:2023nnn,Wang:2021naf,Duan:2021pll},
\begin{eqnarray}
V =\left(\begin{matrix} \frac{{\rho}^0}{\sqrt{2}} + \frac{{\omega}}{\sqrt{2}}  & \rho^+  & K^{*+}  \\
		\rho^-  &   - \frac{{\rho}^0}{\sqrt{2}} + \frac{{\omega}}{\sqrt{2}}  &  K^{*0} \\
		K^{*-}  &  \bar{K}^{*0}   &   \phi
	\end{matrix}
	\right),
\end{eqnarray}
\begin{eqnarray}
	P =\left(\begin{matrix} \frac{\eta}{\sqrt{3}}+ \frac{{\pi}^0}{\sqrt{2}}+ \frac{{\eta}'}{\sqrt{6}} & \pi^+ & K^+  \\
		\pi^-  &   \frac{\eta}{\sqrt{3}}- \frac{{\pi}^0}{\sqrt{2}}+ \frac{{\eta}'}{\sqrt{6}}  &  K^0 \\
		K^-  &  \bar{K}^{0}   &    -\frac{\eta}{\sqrt{3}}+ \frac{{\sqrt{6}\eta}'}{3}
	\end{matrix}
	\right),
\end{eqnarray}
where the $\eta-\eta^{\prime}$ mixing is assumed according to Ref.~\cite{Bramon:1992kr}. The symbol `\textless\textgreater' stands for the  trace of the SU(3) matrices.
One could obtain the relevant contributions by isolating the terms containing $\pi^-$, as follows,
\begin{eqnarray}
&& \textless VVP\textgreater\nonumber\\
&=& (VV)_{12} P_{21}\nonumber\\
&=&\pi^-\sum_i V_{1i}V_{i2} \nonumber\\
    &=&\pi^-\left[\rho^+\left(\frac{\rho^0}{\sqrt{2}}+\frac{\omega}{\sqrt{2}}\right)+\left(-\frac{\rho^0}{\sqrt{2}}+\frac{\omega}{\sqrt{2}}\right)\rho^++\bar{K}^{*0}K^{\ast+}\right] \nonumber\\
    &=&\pi^-\left[\sqrt{2}\rho^+\omega+\bar{K}^{*0}K^{\ast+}\right].\label{eq:3}
\end{eqnarray} 
In the molecular picture, the $a_0(1710)$ is dynamically generated from the $S$-wave $\bar{K}^{*0}K^{\ast+}$ and $\omega\rho^+$ final-state interactions~\cite{Geng:2008gx,Du:2018gyn}, and then decays into the final states $\bar{K}^0K^+$, as depicted in Fig.~\ref{fig:hadronFSI}.
The decay amplitude of Fig.~\ref{fig:hadronFSI} can be written as,
\begin{eqnarray}
\mathcal{M}_a &=& V_p \times \left( G_{\bar{K}^{*0}K^{\ast+}}t_{\bar{K}^{*0}K^{\ast+} \to \bar{K}^0K^+} \right. \nonumber\\
&& \left.+ \sqrt{2}G_{\omega\rho^+}t_{\omega\rho^+ \to \bar{K}^0K^+} \right), \label{eq:ma}
\end{eqnarray}
where $V_P$ is the normalization factor, and $t_{\bar{K}^{*0}K^{\ast+} \to \bar{K}^0K^+}$ and $t_{\omega\rho^+ \to \bar{K}^0K^+}$ are the transition amplitudes.

The loop functions ${G}_{\bar{K}^{*0}K^{\ast+}}$ and ${G}_{\omega\rho^+}$ are  for the $\bar{K}^{*0}{K}^{*+}$  and $\omega\rho^+$ channels, respectively, and read~\cite{Geng:2008gx,Molina:2008jw},
\begin{eqnarray}
	{G}_{i}(M_{\bar{K}^0K^+}) &=& \int_{m_{1-}^2}^{m_{1+}^2} \int_{m_{2-}^2}^{m_{2+}^2}d\tilde{m}_1^2d\tilde{m}_2^2 \times \nonumber \\  
	&& \!\!\!\!\!\!\!\!\!\!\!\!\!\!\!\!\!\!\!\! \omega(\tilde{m}_1^2)\omega(\tilde{m}_2^2)\tilde{G}(M_{\bar{K}^0K^+},\tilde{m}_1^2,\tilde{m}_2^2), \label{eq:loop_vector}
\end{eqnarray}
where
\begin{eqnarray}
\omega(\tilde{m}_i^2) &=& {\frac{1}{N}}\text{Im}\left[\frac{1}{\tilde{m}_i^2-m_{V_{i}}^2+i\Gamma(\tilde{m}_i^2)\tilde{m}_i}\right],
\end{eqnarray}
\begin{eqnarray}
N &=& \int_{\tilde{m}_{i-}^2}^{\tilde{m}_{i+}^2}d\tilde{m}_i^2\text{Im}\left[\frac{1}{\tilde{m}_i^2-m_{V_{i}}^2+i\Gamma(\tilde{m}_i^2)\tilde{m}_i}\right],
\end{eqnarray}
\begin{eqnarray}
\Gamma(\tilde{m}_i^2) = \Gamma_{V_{i}}\frac{\tilde{k}^3}{k^3},
\end{eqnarray}
\begin{align}
	\tilde{k} = \frac{\lambda^\frac{1}{2}(\tilde{m}_i^2,m_{P_{1}}^2,m_{P_{2}}^2)}{2\tilde{m}_i},~	k = \frac{\lambda^\frac{1}{2}(m_{V_i}^2,m_{P_{1}}^2,m_{P_{2}}^2)}{2m_{V_i}},
\end{align}
with the K${\ddot{a}}$llen function $\lambda(x,y,z) =x^2+y^2+z^2-2xy-2xz-2yz$.
Here, we consider the decay channels $\pi\pi$ and $K\pi$ for the vector mesons $\rho$ and $K^*$, respectively, and neglect the contribution from the small width ($\Gamma_\omega=8.68$~MeV) of $\omega$. Taking the vector $K^{\ast}$ for example, $m_{1+}^2=\left(m_{K^\ast}+2\Gamma_{K^\ast}\right)^2$ and $m_{1-}^2=\left(m_{K^\ast}-2\Gamma_{K^\ast}\right)^2$.  Similarly, one can obtain $m_{1+}^2$ and $m_{1-}^2$ for the $\rho$. The masses, widths, and spin-parities of the involved particles are taken from the RPP~\cite{{ParticleDataGroup:2022pth}}, and  listed in Table~\ref{tab:particleparameters}.

\begin{table}[htbp]
\caption{Masses, widths, and spin-parities of the involved particles in this work. All values are in units of MeV.}	\label{tab:particleparameters}
	\begin{tabular}{cccc}
		\hline\hline  
		particle & mass & width  & spin-parity ($J^P$) \\ \hline
		$\eta_c$  &2983.9    &32.0               &$0^-$      \\
		$\pi^\pm$  &139.5704   &---             &$0^-$   \\
		$\bar{K}^0$  &497.611     &---            &$0^-$ \\
        $K^\pm$      &493.677      &---           &$0^-$ \\
         $K^{*}$  &893.6     &49.1          &$1^-$ \\
        $\omega$  &782.65       &8.68           &$1^-$ \\
        $\rho$   &775.26      &149.1           &$1^-$ \\
        $K_0^{\ast}(1430)$ &1425    &270           &$0^+$ \\
        		\hline\hline
	\end{tabular}
\end{table}

The loop function $\tilde{G}$ of Eq.~(\ref{eq:loop_vector}) is for stable particles, and in the dimensional regularization scheme it can be written as~\cite{Geng:2008gx},
\begin{eqnarray}
	\tilde{G} &=& \frac{1}{16\pi^2}\Bigg\{a_{\mu} + \text{ln}\frac{m_1^2}{\mu^2}+\frac{m_2^2-m_1^2+s}{2s}\text{ln}\frac{m_2^2}{m_1^2} \nonumber \\
	&& \frac{p}{\sqrt{s}}\bigg[\text{ln}\left(s-\left(m_2^2-m_1^2\right)+2p\sqrt{s}\right)\nonumber \\
	&& +\text{ln}\left(s+\left(m_2^2-m_1^2\right)+2p\sqrt{s}\right)\nonumber \\
	&& -\text{ln}\left(-s+\left(m_2^2-m_1^2\right)+2p\sqrt{s}\right)\nonumber \\
	&& -\text{ln}\left(-s-\left(m_2^2-m_1^2\right)+2p\sqrt{s}\right)\bigg]\Bigg\},
\end{eqnarray}
with
\begin{eqnarray}
	p=\frac{\lambda^{1/2}(s,m_1^2,m_2^2)}{2\sqrt{s}},
\end{eqnarray}
where $a_{\mu}$ is the subtraction constant, $\mu$ is the dimensional regularization scale, and $s=M^2_{\bar{K}^0K^+}$. We take $a_{\mu} =-1.726$  and $\mu = 1000$~MeV as used in Ref.~\cite{Geng:2008gx}. It is worth mentioning that 
any change in $\mu$ could be reabsorbed by a change in $a_\mu$ through $a_{\mu'} - a_\mu = {\rm ln}(\mu'^{2}/\mu^2)$, which implies that the loop function $\tilde{G}$ is scale independent~\cite{Duan:2022upr}. 

\begin{figure}[htbp]
	\centering
	\includegraphics[scale=0.6]{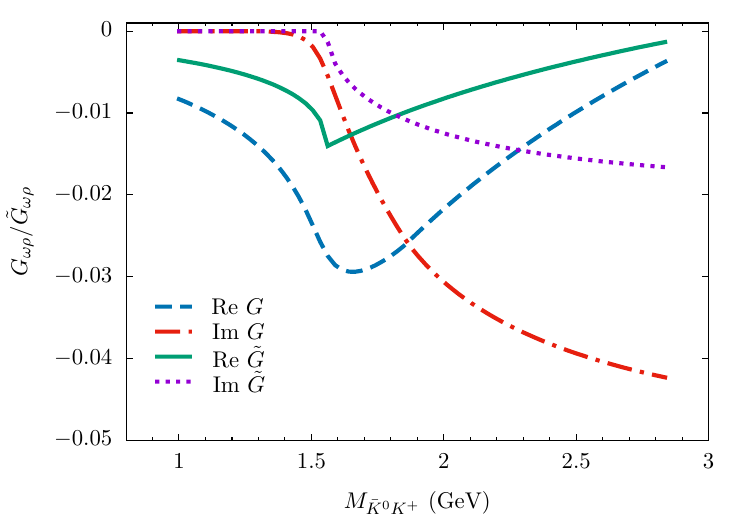}
	\vspace{-0.5cm}
	\caption{Real and imaginary parts of the loop functions $G_{\omega\rho}$ and $\tilde{G}_{\omega\rho}$ as a function of the $\bar{K}^0K^+$ invariant  mass.}
	\label{fig:Gomegarho}
\end{figure}

In order to show the influence of the widths of vector mesons on the loop functions,  we calculate the loop function $G_{\omega\rho}$ and $\tilde{G}_{\omega\rho}$ as  functions of the $\bar{K}^0K^+$ invariant mass, and show them in Fig.~\ref{fig:Gomegarho}. The blue long-dashed and red dot-dashed curves correspond to the real and imaginary parts of the loop function $G$ considering the width of $\rho$, respectively. While, the green solid and purple dotted curves correspond to the real and imaginary parts of the loop function $\tilde{G}$ without the contribution from the $\rho$ width, respectively. One can see that the loop function $G$, considering the width of the vector meson, becomes smoother around the threshold. 

\begin{table}[htbp]
	\caption{Mass, width, and coupling constants of the scalar $a_0(1710)$~\cite{Geng:2008gx}. $g_{K^*{\bar{K}}^*}$, $g_{\omega\rho}$, and $g_{K\bar{K}}$ stand for the coupling constants of $a_0(1710)$ to the $K^*{\bar{K}}^*$,  $\omega\rho$, and $K\bar{K}$ channels, respectively, while  $\Gamma_{K\bar{K}}$ denotes the partial decay width of the $a_0(1710)\to K\bar{K}$. All values are in units of MeV.}	\label{tab:parameters}
	\begin{tabular}{cc}
		\hline\hline  
		 parameters & value\\  \hline \vspace{0.05cm}
	  $M_{a_0}$  &1777  \\ 
	  $\Gamma_{a_0}$ &148.0  \\ 
 $g_{K^*{\bar{K}}^*}$ &(7525, -$i$1529)  \\
 $g_{\omega\rho}$ &(-4042, $i$1393)  \\
$g_{K\bar{K}}$ & 1966  \\
$\Gamma_{K\bar{K}}$ &36    \\
		\hline\hline
	\end{tabular}
\end{table}

On the other hand, the transition amplitudes $t_{\bar{K}^{*0}K^{*+}/\omega\rho^+ \to \bar{K}^0K^+}$ in Eq.~\eqref{eq:ma} could be obtained from the coupled-channel approach in Ref.~\cite{Garcia-Recio:2010enl}, where one state $a_0$ with mass around 1760~MeV could be dynamically generated from the $\eta\pi$, $\bar{K}K$, $\omega\rho$, $\phi\rho$, and $\bar{K}^*K^*$ interactions within SU(6) spin-flavor symmetry. However, the width of $a_0$ is about 24~MeV, much smaller than the one for the $a_0(1710)$ resonance as quoted in the PDG~\cite{ParticleDataGroup:2022pth}. On the other hand, it is customary to obtain the coupling constants and the pole position of the dynamically generated state by fitting the Breit-Wigner form to the amplitude of the coupled-channel approach  around the pole position, 
\begin{equation}
    T_{ij}=\frac{g_ig_j}{s-s_{\rm pole}},
\end{equation}
where $g_{i,j}$ are the couplings to channel $i$ ($j$). It implies that the amplitude of the Breit-Wigner form with the same position and couplings should give similar behavior around the pole position.
Thus, we take the transition amplitude as,
\begin{eqnarray}
t_{\bar{K}^{*0}K^{*+}/\omega\rho^+ \to \bar{K}^0K^+}={\frac{g_{K^*\bar{K}^*/\omega\rho} \times   g_{K\bar{K}}}{M_{\bar{K}^0K^+}^2-M_{a_0}^2+iM_{a_0}\Gamma_{a_0}}},
\end{eqnarray}
where $M_{a_0}$ and $\Gamma_{a_0}$ are the mass and width of the $a_0(1710)$, respectively, and we take their values from Refs.~\cite{Geng:2008gx,Geng:2009gb}, which are tabulated in Table~\ref{tab:parameters}. $g_{K^*{\bar{K}}^*}$, $g_{\omega\rho}$, and $g_{K\bar{K}}$ are the coupling constants of the vertices $K^*{\bar{K}}^*/\omega\rho \to a_0(1710)$\footnote{
The couplings of $a_0(1710)$ to the channels $K^*\bar{K}^*$ and $\omega\rho$
are obtained at the pole position~\cite{Geng:2008gx}. In this work, we take the coupling to be complex, and don't consider the extra phase interference between the coupled-channels $K^*\bar{K}^*$ and $\omega\rho$.} and $a_0(1710)\to K\bar{K}$, respectively, whose values are determined in Ref.~\cite{Geng:2008gx}. We determine the coupling $g_{K\bar{K}}$ from the partial decay width of $a_0(1710) \to K\bar{K}$,
\begin{eqnarray}
\Gamma_{K\bar{K}} = \frac{g^2_{K\bar{K}}}{8\pi} \frac{|{\vec{p}}_K|}{M^2_{a_0}},
\end{eqnarray}
where ${\vec{p}}_K$ is the three momentum of the $K$ or $\bar{K}$ meson in the $a_0(1710)$ rest frame,
\begin{align}
	|{\vec{p}}_K|=\frac{\lambda^{1/2}(M_{a_0}^2,m_{\bar{K}}^2,m_K^2)}{2M_{a_0}}.
\end{align}
With the partial decay width $\Gamma_{K\bar{K}}=36$~MeV~\cite{Geng:2008gx}, one can only obtain the absolute value of the coupling constant, but not the phase, thus we assume that $g_{K\bar{K}}$ is real and positive in this work, as done in Refs.~\cite{Zhu:2022wzk,Zhu:2022guw}.

\subsection{Mechanism of $\eta_c \to ({K^+K_0^{*}(1430)^-}/\bar{K}^0K_0^{*}(1430)^0) \to \bar{K}^0K^+\pi^-$}
\label{Subsec:2B}

\begin{table*}[htbp]
	\centering
	\caption {Coupling constants of the  $K^*_0(1430)$.}	\label{tab:g}
	\begin{tabular}{lcccc}
		\hline\hline  
		Decay process\qquad\quad&Fraction\qquad\quad&Decay width (MeV)\qquad\quad&Coupling constant\qquad\quad& Value (MeV)\\  \hline 
		$\eta_c\to K^+K_0^{\ast-}$\qquad\qquad&$(0.5 \pm 0.1)\%$\qquad\qquad &$32.0\pm0.7$ \qquad\qquad &$g_{\eta_cK^+K_0^{\ast-}}$\qquad\quad&180 \\ $K_0^{\ast}\to K\pi$\qquad\qquad&$(93 \pm 10)\%$\qquad\qquad &$270\pm80$\qquad\qquad& $g_{K_0^{\ast}K\pi}$ \qquad\quad&4721 \\$\eta_c\to\bar{K}^0K_0^{\ast0}$\qquad\qquad&$(0.5 \pm 0.1)\%$\qquad\qquad&$32.0\pm0.7$\qquad\qquad & $g_{\eta_c\bar{K}^0K_0^{\ast0}}$ \qquad\quad&180   \\
		\hline\hline
	\end{tabular}
\end{table*}

\begin{figure}[htbp]
	\centering
	\includegraphics[scale=0.9]{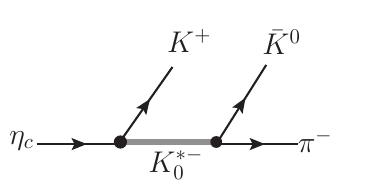}
	\caption{Diagram for $\eta_c\rightarrow  \bar{K}^0K^+\pi^-$ via the intermediate $K_0^{\ast}(1430)^-$, followed by the decay $K_0^{\ast}(1430)^- \to \bar{K}^0\pi^-$.}	\label{fig:mb}
\end{figure}

Firstly, we show the  diagram for the process $\eta_c \to K^+ {K}_0^{*}(1430)^-$, followed by the decay ${{K}}_0^{*}(1430)^-\to \bar{K}^0\pi^-$ in $S$-wave, in Fig.~\ref{fig:mb}.

 The decay amplitude for $\eta_c\rightarrow K^+ K_0^*(1430)^-\to \bar{K}^0K^+\pi^-$ of Fig.~\ref{fig:mb} can be written as
\begin{eqnarray}
    \mathcal{M}_b&=&{\frac{g_{\eta_c K^+ K_0^{\ast-}} g_{K_0^{\ast-}\bar{K}^0\pi^-}} {M_{\bar{K}^0\pi^-}^2-M_{K_0^*}^2+iM_{K_0^*}\Gamma_{K_0^*}}} \label{eq:mb},
\end{eqnarray}
where $M_{\bar{K}^0\pi^-}$ is the invariant mass of the $\bar{K}^0\pi^-$ system, and $g_{\eta_c K^+ K_0^{\ast-}}$ and $g_{K_0^{\ast-}\bar{K}^0\pi^-}$ denote the coupling constants of $\eta_c \to K^+ {K}_0^{*-}$ and ${K}_0^{*-} \to \bar{K}^0\pi^-$, respectively. The mass and width of the $K_0^*(1430)$ are given in Table~\ref{tab:particleparameters}.

\begin{figure}[htbp]
	\centering
	\includegraphics[scale=0.9]{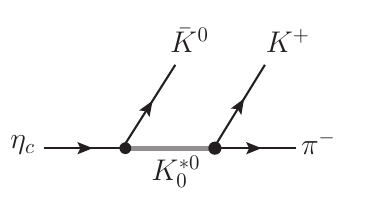}
	\caption{Diagram for $\eta_c\rightarrow \bar{K}^0K^+\pi^-$ via the intermediate $K_0^{\ast}(1430)^0$, followed by the decay $K_0^{\ast}(1430)^0 \to K^+\pi^-$.}	\label{fig:mc}
\end{figure}

Similarly, as shown in Fig.~\ref{fig:mc}, the amplitude of the process $\eta_c \to\bar{K}^0 K_0^*(1430)^0 \to \bar{K}^0K^+\pi^-$ 
 can be expressed as,
\begin{eqnarray}
	\mathcal{M}_c&=&\frac{g_{\eta_c\bar{K}^0K_0^{\ast0}}  g_{K_0^{\ast0}K^+\pi^-}} {M_{K^+\pi^-}^2-M_{K_0^{*}}^2+iM_{K_0^{*}}\Gamma_{K_0^{*}}} \label{eq:g2},
\end{eqnarray}
where $M_{{K}^+\pi^-}$ is the ${K}^+\pi^-$ invariant mass, and $g_{\eta_c \bar{K}^0{K}^{*0}_0}$ and $g_{{K}_0^{*0}K^+\pi^-}$ are the coupling constants of the vertices $\eta_c \to \bar{K}^0 {K}_0^{*}(1430)^0$ and ${K}_0^{*}(1430)^0 \to{K}^+\pi^-$, respectively.

The coupling constants appearing in Eqs.~(\ref{eq:mb}) and (\ref{eq:g2}) could be determined from the experimental partial decay widths of $\eta_c \to K K_0^{*}(1430)$ and $K_0^{*}(1430) \to \bar{K}\pi$, respectively.
The effective Lagrangians accounting for the vertices of $\eta_c\to KK_0^{\ast}(1430)$ and $K_0^{\ast}(1430)\to K\pi$ are given by~\cite{Ding:2008gr},
\begin{eqnarray}
	\mathcal{L}&=&g_{\eta_cKK_0^{\ast}}\eta_cKK_0^{\ast} \label{eq:lag_etac} \\
	\mathcal{L}&=&g_{K_0^*K\pi}K_0^*K\pi. \label{eq:lag_K0}
\end{eqnarray}
With the above effective Lagrangians, we can express the corresponding partial decay widths as, 
\begin{eqnarray}
	\Gamma_{\eta_c \to K_0^*K}&=&\frac{g^2_{\eta_cK_0^{\ast}K}}{8 \pi} \frac{|\mathbf{P}|}{m^2_{\eta_c}}  , \\
	\Gamma_{K_0^* \to K\pi}&=& \frac{{g_{K_0^*K\pi}^2}}{8\pi}\frac{|\mathbf {P}|}{{m^2_{K_0^*}}},
\end{eqnarray}
where $\mathbf{P}$ is the three-momentum of the two final-state particles in the rest frame of the parent particle, which reads,
\begin{align}
	|\mathbf{P}|=\frac{\lambda^{1/2}(M^2,m_1^2,m_2^2)}{2M},
\end{align}
and $M$ and $m_{1,2}$ are the masses of the initial parent particle and the two final-state mesons, respectively. The masses and widths of these particles are given in Table~\ref{tab:particleparameters}.

According to the RPP~\cite{ParticleDataGroup:2022pth}, the branching fraction of $K_0^* \to K\pi$ is $\mathcal{B}(K_0^* \to K\pi) = (93 \pm 10)\%$, and we take it to be $100\% $ in this work. One can then easily obtain the coupling constant $g_{K_0^*K\pi}=4721$~MeV. 

In addition, with the branching fraction $\mathcal{B}(\eta_c \to \bar{K}^0 K^+ \pi^-)=(2.4 \pm 0.2)\% $~\cite{Ji:2021xjz} and the ratio of $\mathcal{B}(\eta_c \to K_0^{\ast0}{\bar{K}}^0/K_0^{\ast-}K^+)/{\mathcal{B}(\eta_c \to {\bar{K}}^0K^+\pi^-)}=(40.8 \pm 2.2)\%$~\cite{BaBar:2015kii} , we could estimate the branching fraction $\mathcal{B}(\eta_c \to K_0^{\ast0}{\bar{K}}^0)=\mathcal{B}(\eta_c \to K_0^{\ast-}K^+)=(0.5 \pm 0.1)\%$. Then, we can determine the coupling constants  $g_{\eta_c K^+ K_0^{*-}}=g_{\eta_c \bar{K}^0 K_0^{*0}}= 180$~MeV.
It is worth mentioning that the coupling constants appearing in Eqs.~(\ref{eq:mb}) and (\ref{eq:g2})  are assumed to be real and positive, and the values of those coupling constants are listed in Table~\ref{tab:g}.

\subsection{Invariant mass distributions}\label{Subsec:2C}

With the amplitudes obtained above, we can write down the total decay amplitude of $\eta_c \to \bar{K}^0K^+\pi^-$ as follows, 
\begin{eqnarray}\label{eq:fullamp}
 \mathcal{M} =\mathcal{M}_a+\mathcal{M}_b+\mathcal{M}_c,
\end{eqnarray}
and the double differential widths of the  process $\eta_c \to \bar{K}^0K^+\pi^-$ are
\begin{eqnarray}
     \frac{d^2\Gamma}{dM_{\bar{K}^0K^+}{dM_{K^+\pi^-}}}=\frac{M_{\bar{K}^0K^+}M_{K^+\pi^-}}{128\pi^3m_{\eta_c}^3}|\mathcal{M}|^2, \label {eq:dgammadm12dm23} \\
     \frac{d^2\Gamma}{dM_{\bar{K}^0K^+}{dM_{\bar{K}^0\pi^-}}}=\frac{M_{\bar{K}^0K^+}M_{\bar{K}^0\pi^-}}{128\pi^3m_{\eta_c}^3}|\mathcal{M}|^2.
\end{eqnarray}
Furthermore, one can easily obtain ${d\Gamma}/{dM_{\bar{K}^0K^+}}$,  ${d\Gamma}/{dM_{\bar{K}^0 {\pi}^-}}$, and ${d\Gamma}/{dM_{K^+{\pi}^-}}$ by integrating over each of the invariant mass variables with the limits of the Dalitz plot given in the RPP~\cite{ParticleDataGroup:2022pth}. For example, the upper and lower limits for $M_{\bar{K}^0K^+}$ are:
\begin{eqnarray}
\left(M_{\bar{K}^0K^+}^2\right)_\text{max} &= &\left(E_{K^+}^\ast+E_{\bar{K}^0}^\ast\right)^2 -  \nonumber \\
    && \left(\sqrt{E_{K^+}^{\ast2}-m_{K^+}^2}-\sqrt{E_{\bar{K}^0}^{\ast2}-m_{\bar{K}^0}^2}\right)^2 \nonumber \\
\left(M_{\bar{K}^0K^+}^2\right)_\text{min} &=&\left(E_{K^+}^\ast+E_{\bar{K}^0}^\ast\right)^2 -  \nonumber \\
    &&\left(\sqrt{E_{K^+}^{\ast2}-m_{K^+}^2}+\sqrt{E_{\bar{K}^0}^{\ast2}-m_{\bar{K}^0}^2}\right)^2, \nonumber
\end{eqnarray}
where $E_{K^+}^\ast$ and $E_{\bar{K}^0}^{\ast}$ are the energies of $K^+$ and $\bar{K}^0$ in the $\bar{K}^0\pi^-$ rest frame, respectively,
\begin{align}
    &E_{\bar{K}^0}^\ast=\frac{M_{\bar{K}^0\pi^-}^2-m_{\pi^-}^2+m_{\bar{K}^0}^2}{2M_{\bar{K}^0\pi^-}},  \nonumber \\
    &E_{K^+}^\ast=\frac{m_{\eta_c}^2-M_{\bar{K}^0\pi^-}^2-m_{K^+}^2}{2M_{\bar{K}^0\pi^-}}.
\end{align}

\section{Results and Discussion} \label{sec:Results}

\begin{figure}[htbp]
	\centering	\includegraphics[scale=0.65]{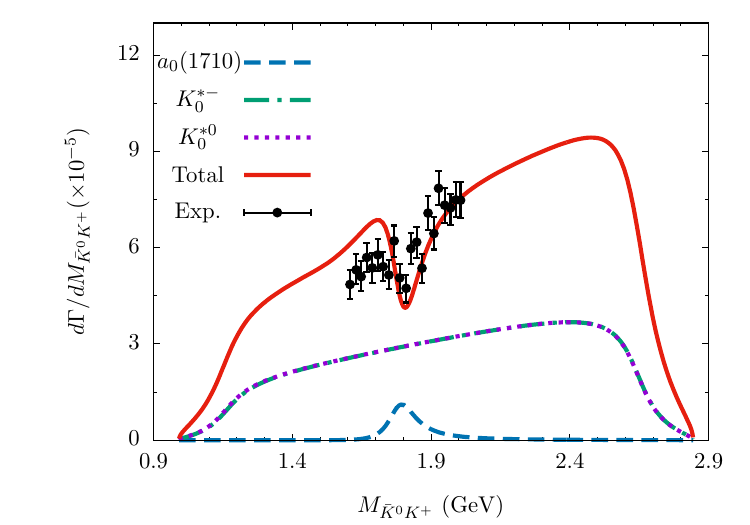}	\caption{$\bar{K}^0K^+$ invariant mass distribution of the process $\eta_c\rightarrow \bar{K}^0K^+\pi^-$. The red-solid curve stands for the total contributions, while the blue-dashed curve, the green-dot-dashed curve, and purple-dotted curve correspond to the contribution from the $a_0(1710)$ state, the intermidiate $K_0^{\ast}(1430)^-$, and $K_0^{*}(1430)^0$, respectively. The {\it BABAR} data are taken from Fig. 7(a) of Ref.~\cite{BaBar:2015kii}. }	\label{fig:ks0kz}
\end{figure}

\begin{figure}[htbp]
	\centering	\includegraphics[scale=0.65]{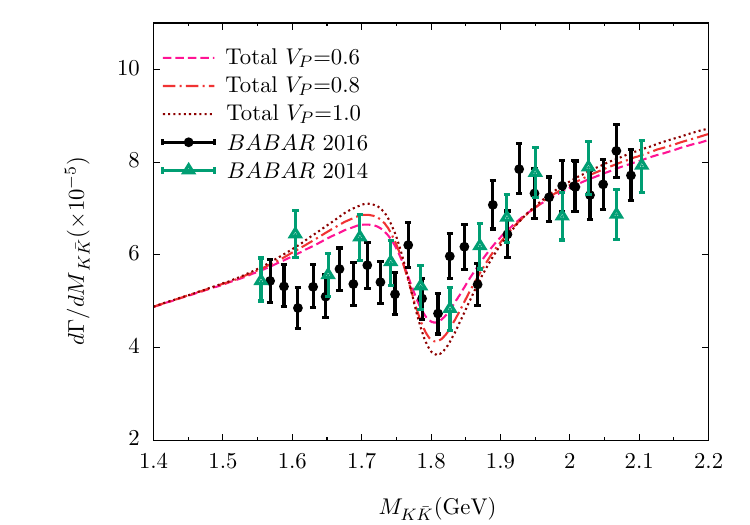}
	\caption{$\bar{K}^0K^+$ invariant mass distribution of the process $\eta_c\rightarrow \bar{K}^0K^+\pi^-$ with the parameter $V_p=0.6$, $0.8$, $1.0$, respectively. In addition to the {\it BABAR} measurements of the $\eta_c\rightarrow \bar{K}^0K^+\pi^-$~\cite{BaBar:2015kii} (labeled as `{\it BABAR} 2016'), we also show the {\it BABAR} measurements of the $K^+K^-$ invariant mass distribution of the process $\eta_c\to K^+K^-\pi^0$~\cite{BaBar:2014asx} (labeled as `{\it BABAR} 2014'). }	\label{fig:vp}
\end{figure}

\begin{figure}[htbp]
	\centering	\includegraphics[scale=0.65]{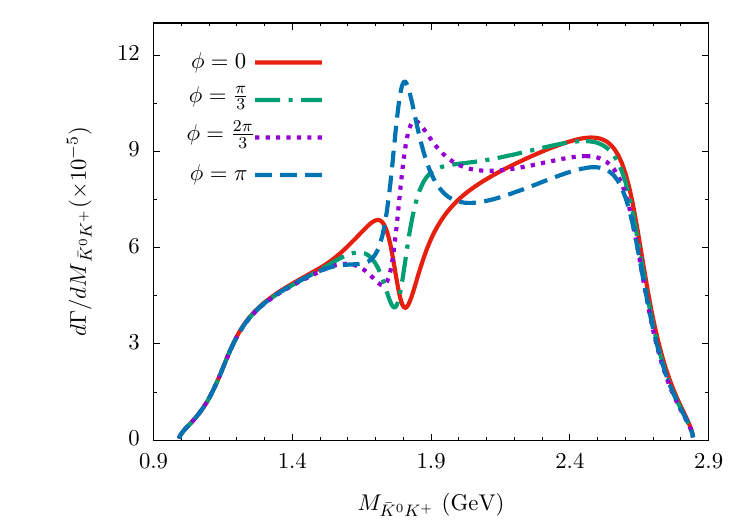}
	\caption{$\bar{K}^0K^+$ invariant mass distribution of the process $\eta_c\rightarrow \bar{K}^0K^+\pi^-$ obtained with a phase angle $\phi=0$, $\pi/3$, $2\pi/3$, and $\pi$, respectively. See the text for details.  }	\label{fig:phase}
\end{figure}

\begin{figure}[htbp]
	\centering
	\includegraphics[scale=0.65]{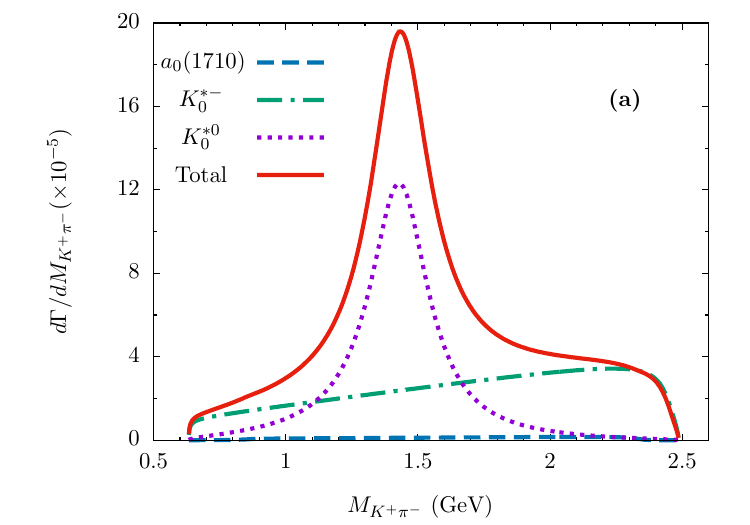}
 \includegraphics[scale=0.65]{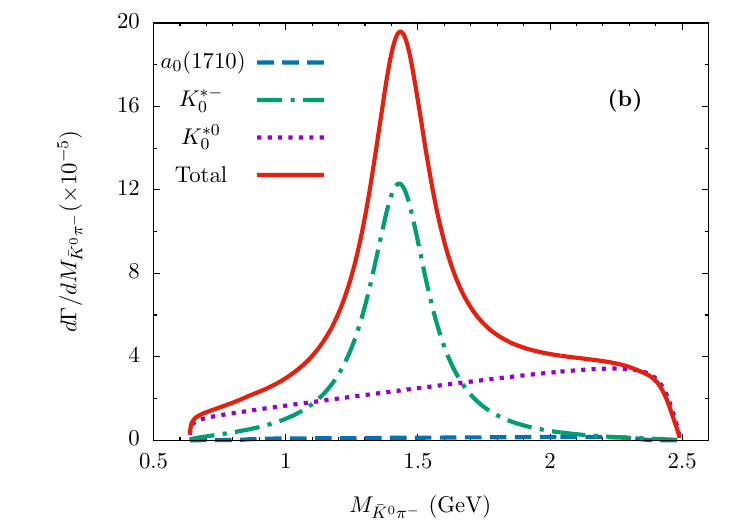}
	\caption{$K^+\pi^-$ (a) and $\bar{K}^0\pi^-$ (b) invariant mass distribution of the process $\eta_c\rightarrow \bar{K}^0K^+\pi^-$. The notations of the curves are the same as those of Fig.~\ref{fig:ks0kz}.}	\label{fig:dw_Kpi}
\end{figure}

\begin{figure}[htbp]
	\includegraphics[scale=0.75]{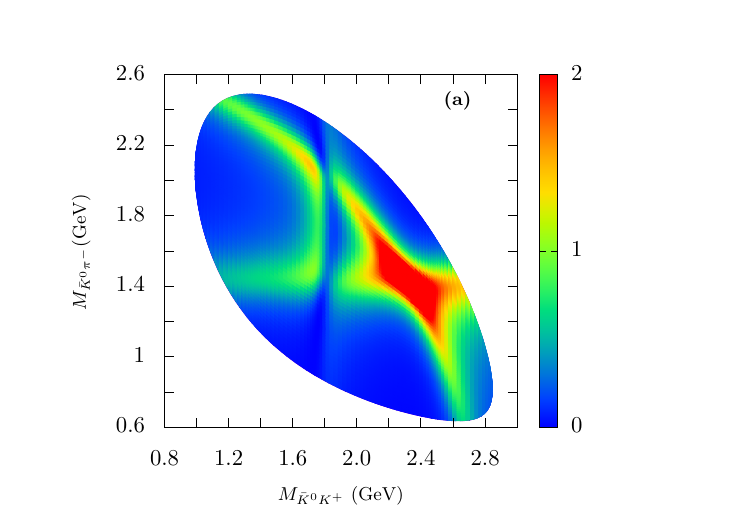} 
	\includegraphics[scale=0.75]{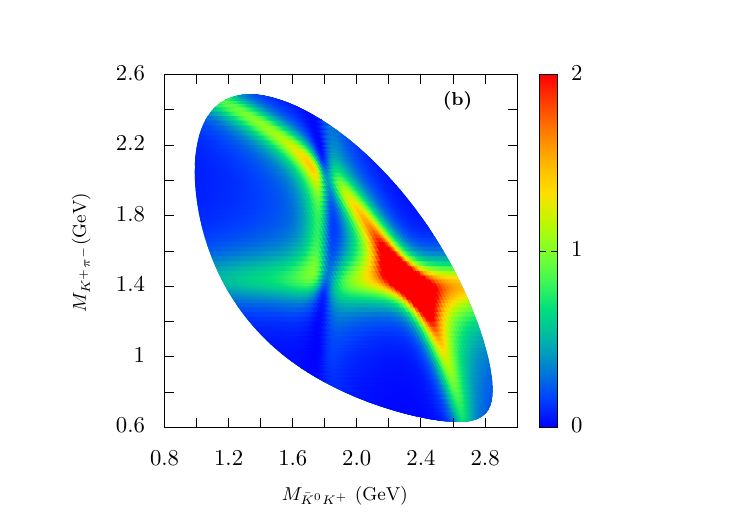} 
		\includegraphics[scale=0.75]{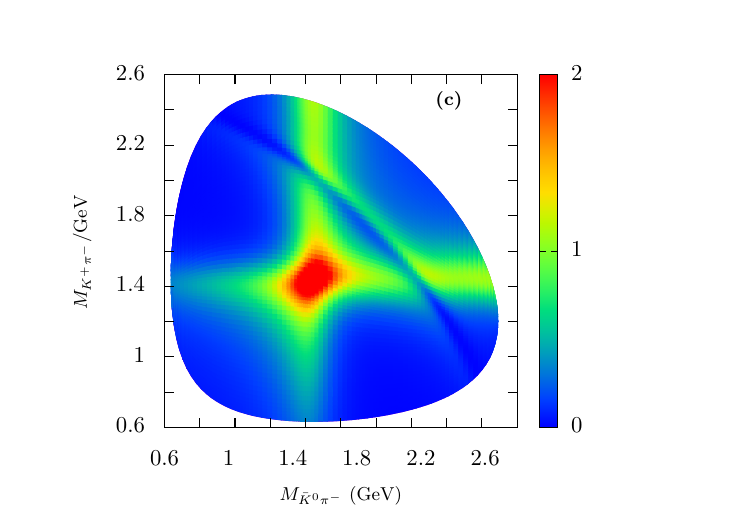} 
	\caption{Dalitz plots for the decay $\eta_c\rightarrow \bar{K}^0K^+\pi^-$.  (a) $M_{\bar{K}^0K^+}$ vs. $M_{\bar{K}^0\pi^-}$; (b) $M_{\bar{K}^0K^+}$ vs. $M_{K^+\pi^-}$; (c) $M_{\bar{K}^0K^+}$ vs. $M_{K^+\pi^-}$. }	\label{fig:dalitz}
\end{figure}
 It should be pointed out that the $K_S^0K^+$ invariant mass distribution of the process  $\eta_c\rightarrow {K}^0_SK^+\pi^-$ has been measured by the {\it BABAR} Collaboration~\cite{BaBar:2015kii}. In this work, we take $V_p=0.8$ in order to match with the {\it BABAR} measurements of the $K_S^0K^+$ invariant mass distribution around $1.6\sim 2.1$~GeV. In Fig.~\ref{fig:ks0kz}, we show our results of the  $\bar{K}^0K^+$ invariant mass distribution, where the red-solid curve stands for the total contribution from the $a_0(1710)$ state and the vector $K_0^*$ meson, while the blue-dashed curve corresponds to the contribution from  the $a_0(1710)$ state. Moreover, the green-dot-dashed and purple-dotted  curves stand for the contributions from the intermediate $K_0^{\ast}(1430)^-$ and $K_0^{*}(1430)^0$, respectively. We also show the {\it BABAR} data points in the region of 1.6 $\sim$ 2.1~GeV\footnote{As pointed out in Ref.~\cite{BaBar:2015kii},  for the $\eta_c\to\bar{K}^0K^+\pi^-$ decay, some other resonances also contribute, such as the $a_0(980)$, $a_0(1450)$, $a_0(1950)$, and $a_2(1320)$. Since in this work we focus on the possible signal of the $a_0(1710)$, only the {\it BABAR} data in the region of $1.6 \sim 2.1$~GeV are presented in Figs.~\ref{fig:ks0kz} and \ref{fig:vp}.}, which has been multiplied by an overall normalization factor $4\times10^{-7}$~\cite{BaBar:2015kii}. As one can see from Fig.~\ref{fig:ks0kz}, the contributions from the $K_0^*(1430)$ are smooth in the region of $1.4\sim 2.4$~GeV. In particular, we note that the dip structure around 1800~MeV is in agreement with the {\it BABAR} measurement~\cite{BaBar:2015kii}. This dip structure is mainly due to the interference between the contributions from the $a_0(1710)$ and the $K_0^{\ast}(1430)$, and should be associated to the scalar $a_0(1710)$. 

In order to show the dependence of our results on the parameter $V_p$, we present the $\bar{K}^0K^+$ invariant mass distribution of the process $\eta_c\rightarrow \bar{K}^0K^+\pi^-$ with the parameter $V_p=0.6, 0.8, 1.0$ in Fig.~\ref{fig:vp}.  One can see that the dip structure around 1.8~GeV persists, which is  in agreement with the {\it BABAR} measurements~\cite{BaBar:2015kii}, labeled as `{\it BABAR} 2016'.  It should be stressed that the {\it BABAR} Collaboration has also measured the $K^+K^-$ invariant mass distribution of the process $\eta_c\to K^+K^-\pi^0$, as shown by the data of `{\it BABAR} 2014' in Fig.~\ref{fig:vp},  where one dip structure also appears around 1.8~GeV~\cite{BaBar:2014asx}.

However, it should be pointed out that the dip structure appearing in the $\bar{K}^0K^+$ invariant mass distribution of Fig.~\ref{fig:ks0kz} could also manifest itself 
as a peak structure if the interference between $\mathcal{M}_a$, $\mathcal{M}_b$, and 
$\mathcal{M}_c$ are different from our naive assignments explained above. For instance, if we multiply the term $\mathcal{M}_a$ of
Eq.~(\ref{eq:fullamp}) by a phase factor $e^{i\phi}$ with $\phi=0$, $\pi/3$, $2\pi/3$, and $\pi$, we would obtain the $\bar{K}^0K^+$ invariant mass distributions shown in Fig.~\ref{fig:phase}, where one can see a peak structure around 1.8~GeV  for $\phi=2\pi/3$ and $\pi$.

Next, with the parameter $V_p=0.8$, we predict the $K^+\pi^-$ and $\bar{K}^0\pi^-$ invariant mass distributions for the  $\eta_c\rightarrow \bar{K}^0K^+\pi^-$ in Figs.~\ref{fig:dw_Kpi}(a) and (b), respectively. One can see the clear peaks of the $K_0^*(1430)^0$ and $K_0^*(1430)^-$ , which is consistent with the {\it BABAR} measurements [see Figs.~5(a) and 5(b) of Ref.~\cite{BaBar:2015kii}].

In Fig.~\ref{fig:dalitz}, we present the Dalitz plots for the process $\eta_c\rightarrow \bar{K}^0K^+\pi^-$ with the parameter $V_p=0.8$. From Figs.~\ref{fig:dalitz}(a) and \ref{fig:dalitz}(b), we can clearly see that there is a vertical blue band around $M_{\bar{K}^0K^+}=1.8$~GeV, which should be associated with the signal of the scalar $a_0(1710)$, and we also find a yellow band around $M_{\bar{K}^0\pi^-/K^+\pi^-}=1.43$~GeV, corresponding to the signal of the  ${K_0^*(1430)}$ state. From Fig.~\ref{fig:dalitz}(c), can see that  most events of the process $\eta_c\rightarrow \bar{K}^0K^+\pi^-$ will appear in the region around $M_{\bar{K}^0\pi^-}=1.43$~GeV and $M_{K^+\pi^-}=1.43$~GeV, which is in agreement with the {\it BABAR} measurements (see Fig.~4 of Ref.~\cite{BaBar:2015kii}).

Finally, we predict the branching fractions of the processes $\eta_c \to \bar{K}^{*0}K^{\ast+}\pi^-$ and $\eta_c \to \omega\rho^+\pi^-$, which have not yet been measured. Without the contributions from intermediate resonances, based on Eq.~(\ref{eq:3}) the amplitudes for the processes  $\eta_c \to \bar{K}^{*0}K^{\ast+}\pi^-$ and $\eta_c \to \omega\rho^+\pi^-$ are,
\begin{eqnarray}
\mathcal{M}^{\eta_c \to \bar{K}^{*0}K^{\ast+}\pi^-}&=& V_p \vec\epsilon_{\bar{K}^{*0}}\cdot \vec\epsilon_{K^{\ast+}}, \\
\mathcal{M}^{\eta_c \to \omega\rho^+\pi^-}&=&\sqrt{2} V_p \vec\epsilon_{ \omega}\cdot \vec\epsilon_{\rho^+}, 
\end{eqnarray}
where $\vec{\epsilon}_i$ is the polarization of the vector meson, and $\sum_{\rm pol}\epsilon_i(R)\epsilon^*_j(R)=\delta_{ij}$~\cite{Duan:2023qsg}.  With the parameter $V_p=0.8$, we could  estimate the branching fractions of these two processes,
\begin{eqnarray}
 \mathcal{B}(\eta_c \to \bar{K}^{*0}K^{\ast+}\pi^-)&=&\frac{1}{\Gamma_{\eta_c}}\int\left(\frac{d\Gamma}{dM_{\bar{K}^{*0}K^{\ast+}}}\right)dM_{\bar{K}^{*0}K^{\ast+}} \nonumber  \\
 &=&5.5\times 10^{-3} \\
  \mathcal{B}(\eta_c \to\omega\rho^+\pi^-)&=& \frac{1}{\Gamma_{\eta_c}}  \int\left(\frac{d\Gamma}{dM_{\omega\rho^+}}\right)dM_{\omega\rho^+} \nonumber \\
  &=&7.9\times10^{-3} ,
\end{eqnarray}
where the formalism of the differential width of the three-body decay could be found in the RPP~\cite{ParticleDataGroup:2022pth}. We note that our prediction for $\mathcal{B}(\eta_c \to \bar{K}^{*0}K^{\ast+}\pi^-)=5.5\times 10^{-3}$ is less than $\mathcal{B}(\eta_c \to K^+K^-\pi^+\pi^-\pi^0)=(3.4\pm0.5) \%$ and  $\mathcal{B}(\eta_c \to K^0K^-\pi^+\pi^-\pi^+)=(5.7\pm 1.6) \%$, while the prediction for $ \mathcal{B}(\eta_c \to\omega\rho^+\pi^-)=7.9\times 10^{-3}$ is less than $\mathcal{B}(\eta_c \to 2(\pi^+\pi^-\pi^0)=(16.2\pm 2.1) \%$~\cite{ParticleDataGroup:2022pth}, which seem  reasonable.

The BESIII Collaboration has collected 10 billion $J/\psi$ events and 3 billion $\psi(3686)$ events, and the available $\eta_c$ events via the decays of $J/\psi\to \gamma \eta_c$ and $\psi(3686)\to \gamma \eta_c$ are recently proposed to precisely measure the $\eta_c$ decay modes~\cite{Ji:2021xjz}, which could be helpful to search for the possible signal of the $a_0(1710)$, and test our theoretical predictions.
The $\eta_c \to \bar{K}^0 K^+\pi^-$ reaction could be a good platform to investigate the $a_0(1710)$, especially  its mass.

It should be stressed that one can not exclude the other interpretations based the present experimental information. In Ref.~\cite{Achasov:2023izs}, the authors have studied the coupled-channels influence on the $a_0(1710)$ line shape by assuming it as four-quark state in the MIT bag model, and found that the strong couplings of $a_0$ to $VV$ channel can narrow the $a_0$ peak in the $PP$ mass spectra, and the $a_0$ width could be $150\sim 300$~MeV in the absence of $K\bar{K}$ and $\pi\eta$ channels. It is suggested to detect the $a_0(1710)\to VV$ decay directly to test their results in Ref.~\cite{Achasov:2023izs}.

\section{Summary} \label{sec:Conclusions}
Assuming the $a_0(1710)$ as a $K^*\bar{K}^*$ molecular state, we have investigated the process $\eta_c\rightarrow \bar{K}^0K^+\pi^-$ taking into account the contribution from the $S$-wave $\omega\rho^+$ and $\bar K^{*0}K^{*+}$ interactions, as well as the contribution from the intermediate resonance $K^*_0(1430)$. 
We predicted one dip structure around 1.8~GeV in the $\bar{K}^0K^+$ invariant mass distribution, which is in agreement with the {\it BABAR} measurements~\cite{BaBar:2015kii}. It should be pointed out that a similar dip structure also appears around 1.8~GeV in the $K^+K^-$ invariant mass distribution of the process $\eta_c\to K^+K^-\pi^0$ of the {\it BABAR} measurements~\cite{BaBar:2014asx}. Furthermore, we predicted the $K^+\pi^-$ and $\bar{K}^0\pi^-$ invariant mass distributions of the process $\eta_c\rightarrow \bar{K}^0K^+\pi^-$, and found clear peaks of the resonance $K_0^*(1430)^{0,-}$, consistent with the {\it BABAR} measurements~\cite{BaBar:2015kii}. In addition, we have also plotted the Dalitz plots of the process $\eta_c\rightarrow \bar{K}^0K^+\pi^-$, and shown the possible signals of the $a_0(1710)$ and $K^*_0(1430)$.

Finally, we have estimated the branching fractions $\mathcal{B}(\eta_c \to \bar{K}^{*0}K^{\ast+}\pi^-)=5.5\times 10^{-3}$ and $ \mathcal{B}(\eta_c \to\omega\rho^+\pi^-)=7.9\times 10^{-3}$, which are reasonable by comparing with the experimental data. Our theoretical predictions could be tested by the BESIII and Belle II experiments in the future, and the precise measurements of the process $\eta_c\rightarrow \bar{K}^0K^+\pi^-$ could  shed light on the nature of the scalar $a_0(1710)$.

\section*{Acknowledgments}

We would like to thank Profs. Wen-Cheng Yan and Ya-Teng Zhang for useful discussions. This work is partly supported by the National Natural
Science Foundation of China under Grants Nos. 12075288, 11975041,  11961141004, 11961141012, and 12192263.
This work is supported by the Natural Science Foundation of Henan under Grant Nos. 222300420554, 232300421140, the Project of Youth Backbone Teachers of Colleges and Universities of Henan Province (2020GGJS017), and the Open Project of Guangxi Key Laboratory of Nuclear Physics and Nuclear Technology, No. NLK2021-08.


\begin{thebibliography}{10}
	
	
\bibitem{BaBar:2021fkz}
J.~P.~Lees \textit{et al.} [BaBar],
Phys. Rev. D \textbf{104} (2021) no.7, 072002
doi:10.1103/PhysRevD.104.072002
[arXiv:2106.05157 [hep-ex]].
	
	
\bibitem{BESIII:2021anf}
M.~Ablikim \textit{et al.} [BESIII],
Phys. Rev. D \textbf{105} (2022) no.5, L051103
doi:10.1103/PhysRevD.105.L051103
[arXiv:2110.07650 [hep-ex]].
	
\bibitem{BESIII:2022npc}
M.~Ablikim \textit{et al.} [BESIII],
Phys. Rev. Lett. \textbf{129} (2022) no.18, 182001
doi:10.1103/PhysRevLett.129.182001
[arXiv:2204.09614 [hep-ex]].
	
\bibitem{Nagahiro:2008bn}
H.~Nagahiro, L.~Roca, E.~Oset and B.~S.~Zou,
Phys. Rev. D \textbf{78} (2008), 014012
doi:10.1103/PhysRevD.78.014012
[arXiv:0803.4460 [hep-ph]].
	
\bibitem{Branz:2009cv}
T.~Branz, L.~S.~Geng and E.~Oset,
Phys. Rev. D \textbf{81} (2010), 054037
doi:10.1103/PhysRevD.81.054037
[arXiv:0911.0206 [hep-ph]].
	
\bibitem{Geng:2010kma}
L.~S.~Geng, F.~K.~Guo, C.~Hanhart, R.~Molina, E.~Oset and B.~S.~Zou,
Eur. Phys. J. A \textbf{44} (2010), 305-311
doi:10.1140/epja/i2010-10971-5
[arXiv:0910.5192 [hep-ph]].
	
\bibitem{Xie:2014gla}
J.~J.~Xie and E.~Oset,
Phys. Rev. D \textbf{90} (2014) no.9, 094006
doi:10.1103/PhysRevD.90.094006
[arXiv:1409.1341 [hep-ph]].
	
\bibitem{MartinezTorres:2012du}
A.~Martinez Torres, K.~P.~Khemchandani, F.~S.~Navarra, M.~Nielsen and E.~Oset,
Phys. Lett. B \textbf{719} (2013), 388-393
doi:10.1016/j.physletb.2013.01.036
[arXiv:1210.6392 [hep-ph]].
	
\bibitem{Wang:2021jub}
Z.~L.~Wang and B.~S.~Zou,
Phys. Rev. D \textbf{104} (2021) no.11, 114001
doi:10.1103/PhysRevD.104.114001
[arXiv:2107.14470 [hep-ph]].
	
\bibitem{Garcia-Recio:2010enl}
C.~Garcia-Recio, L.~S.~Geng, J.~Nieves and L.~L.~Salcedo,
Phys. Rev. D \textbf{83} (2011), 016007
doi:10.1103/PhysRevD.83.016007
[arXiv:1005.0956 [hep-ph]].


\bibitem{Garcia-Recio:2013uva}
C.~Garc\'\i{}a-Recio, L.~S.~Geng, J.~Nieves, L.~L.~Salcedo, E.~Wang and J.~J.~Xie,
Phys. Rev. D \textbf{87} (2013) no.9, 096006
doi:10.1103/PhysRevD.87.096006
[arXiv:1304.1021 [hep-ph]].


	
\bibitem{Close:2005vf}
F.~E.~Close and Q.~Zhao,
Phys. Rev. D \textbf{71} (2005), 094022
doi:10.1103/PhysRevD.71.094022
[arXiv:hep-ph/0504043 [hep-ph]].

\bibitem{Gui:2012gx}
L.~C.~Gui \textit{et al.} [CLQCD],
Scalar Glueball in Radiative $J/\psi$ Decay on the Lattice,
Phys. Rev. Lett. \textbf{110}, 021601 (2013).

\bibitem{Janowski:2014ppa}
S.~Janowski, F.~Giacosa and D.~H.~Rischke,
Is $f_0(1710)$ a glueball?,
Phys. Rev. D \textbf{90}, 114005 (2014).

\bibitem{Fariborz:2015dou}
A.~H.~Fariborz, A.~Azizi and A.~Asrar,
Proximity of $f_0$(1500) and $f_0$(1710) to the scalar glueball,
Phys. Rev. D \textbf{92}, 113003 (2015).

\bibitem{Chen:2005mg}
Y.~Chen, A.~Alexandru, S.~J.~Dong, T.~Draper, I.~Horvath, F.~X.~Lee, K.~F.~Liu, N.~Mathur, C.~Morningstar and M.~Peardon, \textit{et al.}
Phys. Rev. D \textbf{73} (2006), 014516
doi:10.1103/PhysRevD.73.014516
[arXiv:hep-lat/0510074 [hep-lat]].

\bibitem{ParticleDataGroup:2022pth}
R.~L.~Workman \textit{et al.} [Particle Data Group],
PTEP \textbf{2022} (2022), 083C01
doi:10.1093/ptep/ptac097


\bibitem{Geng:2008gx}
L.~S.~Geng and E.~Oset,
Phys. Rev. D \textbf{79} (2009), 074009
doi:10.1103/PhysRevD.79.074009
[arXiv:0812.1199 [hep-ph]].

\bibitem{Du:2018gyn}
M.~L.~Du, D.~G\"ulmez, F.~K.~Guo, U.~G.~Mei\ss{}ner and Q.~Wang,
Eur. Phys. J. C \textbf{78} (2018) no.12, 988
doi:10.1140/epjc/s10052-018-6475-8
[arXiv:1808.09664 [hep-ph]].


\bibitem{Wang:2022pin}
Z.~L.~Wang and B.~S.~Zou,
Eur. Phys. J. C \textbf{82} (2022) no.6, 509
doi:10.1140/epjc/s10052-022-10460-4
[arXiv:2203.02899 [hep-ph]].

\bibitem{Wang:2017pxm}
G.~Y.~Wang, S.~C.~Xue, G.~N.~Li, E.~Wang and D.~M.~Li,
Phys. Rev. D \textbf{97} (2018) no.3, 034030
doi:10.1103/PhysRevD.97.034030
[arXiv:1712.10180 [hep-ph]].

	
\bibitem{Guo:2022xqu}
D.~Guo, W.~Chen, H.~X.~Chen, X.~Liu and S.~L.~Zhu,
Phys. Rev. D \textbf{105} (2022) no.11, 114014
doi:10.1103/PhysRevD.105.114014
[arXiv:2204.13092 [hep-ph]].
	
\bibitem{BES:2006vdb}
M.~Ablikim \textit{et al.} [BES],
Phys. Rev. Lett. \textbf{96} (2006), 162002
doi:10.1103/PhysRevLett.96.162002
[arXiv:hep-ex/0602031 [hep-ex]].

\bibitem{BESIII:2012rtd}
M.~Ablikim \textit{et al.} [BESIII],
Phys. Rev. D \textbf{87} (2013) no.3, 032008
doi:10.1103/PhysRevD.87.032008
[arXiv:1211.5668 [hep-ex]].
	

\bibitem{Zhu:2022wzk}
X.~Zhu, D.~M.~Li, E.~Wang, L.~S.~Geng and J.~J.~Xie,
Phys. Rev. D \textbf{105} (2022) no.11, 116010
doi:10.1103/PhysRevD.105.116010
[arXiv:2204.09384 [hep-ph]].
	
\bibitem{Zhu:2022guw}
X.~Zhu, H.~N.~Wang, D.~M.~Li, E.~Wang, L.~S.~Geng and J.~J.~Xie,
Phys. Rev. D \textbf{107} (2023) no.3, 034001
doi:10.1103/PhysRevD.107.034001
[arXiv:2210.12992 [hep-ph]].
	

\bibitem{Dai:2021owu}
L.~R.~Dai, E.~Oset and L.~S.~Geng,
Eur. Phys. J. C \textbf{82} (2022) no.3, 225
doi:10.1140/epjc/s10052-022-10178-3
[arXiv:2111.10230 [hep-ph]].

\bibitem{Oset:2023hyt}
E.~Oset, L.~R.~Dai and L.~S.~Geng,
Sci. Bull. \textbf{68} (2023), 243-246
doi:10.1016/j.scib.2023.01.011
[arXiv:2301.08532 [hep-ph]].

\bibitem{Wang:2023aza}
Z.~Y.~Wang, Y.~W.~Peng, J.~Y.~Yi, W.~C.~Luo and C.~W.~Xiao,
Phys. Rev. D \textbf{107} (2023) no.11, 116018
doi:10.1103/PhysRevD.107.116018

\bibitem{Wang:2023lia}
X.~Y.~Wang, H.~F.~Zhou and X.~Liu,
Phys. Rev. D \textbf{108} (2023) no.3, 034015
doi:10.1103/PhysRevD.108.034015
[arXiv:2306.12815 [hep-ph]].


\bibitem{BESIII:2012urf}
M.~Ablikim \textit{et al.} [BESIII],
Phys. Rev. D \textbf{86} (2012), 092009
doi:10.1103/PhysRevD.86.092009
[arXiv:1209.4963 [hep-ex]].

\bibitem{BESIII:2019eyx}
M.~Ablikim \textit{et al.} [BESIII],
Phys. Rev. D \textbf{100} (2019) no.1, 012003
doi:10.1103/PhysRevD.100.012003
[arXiv:1903.05375 [hep-ex]].



	
\bibitem{BaBar:2015kii}
J.~P.~Lees \textit{et al.} [BaBar],
Phys. Rev. D \textbf{93} (2016), 012005
doi:10.1103/PhysRevD.93.012005
[arXiv:1511.02310 [hep-ex]].
	
\bibitem{BaBar:2010siw}
J.~P.~Lees \textit{et al.} [BaBar],
Phys. Rev. D \textbf{81} (2010), 052010
doi:10.1103/PhysRevD.81.052010
[arXiv:1002.3000 [hep-ex]].
	
	
	
\bibitem{Ikeno:2019grj}
N.~Ikeno, J.~M.~Dias, W.~H.~Liang and E.~Oset,
Phys. Rev. D \textbf{100} (2019) no.11, 114011
doi:10.1103/PhysRevD.100.114011
[arXiv:1909.11906 [hep-ph]].
	

	

\bibitem{Jiang:2019ijx}
S.~J.~Jiang, S.~Sakai, W.~H.~Liang and E.~Oset,
Phys. Lett. B \textbf{797} (2019), 134831
doi:10.1016/j.physletb.2019.134831
[arXiv:1904.08271 [hep-ph]].


\bibitem{Zhang:2023nnn}
H.~Zhang, B.~C.~Ke, Y.~Yu and E.~Wang,
Chin. Phys. C \textbf{47} (2023) no.6, 063101
doi:10.1088/1674-1137/acc642
[arXiv:2302.10541 [hep-ph]].

\bibitem{Wang:2021naf}
J.~Y.~Wang, M.~Y.~Duan, G.~Y.~Wang, D.~M.~Li, L.~J.~Liu and E.~Wang,
Phys. Lett. B \textbf{821} (2021), 136617
doi:10.1016/j.physletb.2021.136617
[arXiv:2105.04907 [hep-ph]].
	
\bibitem{Duan:2021pll}
M.~Y.~Duan, G.~Y.~Wang, E.~Wang, D.~M.~Li and D.~Y.~Chen,
Phys. Rev. D \textbf{104} (2021) no.7, 074030
doi:10.1103/PhysRevD.104.074030
[arXiv:2109.00731 [hep-ph]].
	
\bibitem{Bramon:1992kr}
A.~Bramon, A.~Grau and G.~Pancheri,
Phys. Lett. B \textbf{283} (1992), 416-420
doi:10.1016/0370-2693(92)90041-2
	

	
	
\bibitem{Molina:2008jw}
R.~Molina, D.~Nicmorus and E.~Oset,
Phys. Rev. D \textbf{78} (2008), 114018
doi:10.1103/PhysRevD.78.114018
[arXiv:0809.2233 [hep-ph]].
	

\bibitem{Duan:2022upr}
M.~Y.~Duan, D.~Y.~Chen and E.~Wang,
Eur. Phys. J. C \textbf{82} (2022) no.10, 968
doi:10.1140/epjc/s10052-022-10948-z
[arXiv:2207.03930 [hep-ph]].
	



\bibitem{Geng:2009gb}
L.~S.~Geng, E.~Oset, R.~Molina and D.~Nicmorus,
PoS \textbf{EFT09} (2009), 040
doi:10.22323/1.069.0040
[arXiv:0905.0419 [hep-ph]].
	
	
\bibitem{Ding:2008gr}
G.~J.~Ding,
Phys. Rev. D \textbf{79} (2009), 014001
doi:10.1103/PhysRevD.79.014001
[arXiv:0809.4818 [hep-ph]].

	
\bibitem{Ji:2021xjz}
Q.~Ji, S.~Fang and Z.~Wang,
[arXiv:2108.13029 [hep-ex]].
	
		
\bibitem{BaBar:2014asx}
J.~P.~Lees \textit{et al.} [BaBar],
Phys. Rev. D \textbf{89} (2014) no.11, 112004
doi:10.1103/PhysRevD.89.112004
[arXiv:1403.7051 [hep-ex]].




\bibitem{Duan:2023qsg}
M.~Y.~Duan, E.~Wang and D.~Y.~Chen,
[arXiv:2305.09436 [hep-ph]].

\bibitem{Achasov:2023izs}
N.~N.~Achasov and G.~N.~Shestakov,
Phys. Rev. D \textbf{108} (2023) no.3, 036018
doi:10.1103/PhysRevD.108.036018
[arXiv:2306.04478 [hep-ph]].
	
\end{thebibliography}
\end{document}